\documentclass{article}
\usepackage{graphicx} % Required for inserting images
\usepackage{amsmath}  % For better math formatting
\usepackage{longtable} % For long tables
\usepackage{dutchcal}  % For cool math caligraphic fonts
\usepackage{amssymb}  % For \mathbb and extra math symbols
\usepackage[margin=1in]{geometry}
\usepackage{enumitem}
\usepackage{caption}
\usepackage{float}
\usepackage{booktabs}
\usepackage{siunitx} % For S column type in tables
\usepackage{titlesec}
\usepackage{hyperref}

\title{The Price of Disaster: Estimating the Impact of Hurricane Harvey on the Texas Construction Labor Market}
\author{
  Kartik Ganesh\thanks{Institute for Computing in Research}
}
\date{July 2025}

\begin{document}

\maketitle

\begin{abstract}
This paper estimates the effect of Hurricane Harvey on wages and employment in the construction labor industry across impacted counties in Texas. Based on data from the Quarterly Census of Employment and Wages (QCEW) for the period 2016-2019, I adopted a difference-in-differences event study approach by comparing results in 41 FEMA-designated disaster counties with a set of unaffected southern control counties. I find that Hurricane Harvey had a large and long-lasting impact on labor market outcomes in the construction industry. More precisely, average log wages in treated counties rose by around 7.2 percent compared to control counties two quarters after the hurricane and remained high for the next two years. Employment effects were more gradual, showing a statistically significant increase only after six quarters, in line with the lagged nature of large-scale reconstruction activities. These results imply that natural disasters can generate persistent labor demand shocks to local construction markets, with policy implications for disaster recovery planning and workforce mobilization.
\end{abstract}

\section{Introduction}
Natural disasters impose serious and often long-lasting changes on local labor markets [1, 2]. Previous studies show that hurricanes can lead to wage and employment volatility [1,4]. Understanding how local economies respond to these shocks is essential for designing effective disaster recovery policies. In particular, the construction sector plays an essential role in rebuilding efforts and often experiences unique labor market dynamics in the aftermath of a disaster.

Hurricane Harvey, one of the most destructive storms in the history of the United States of America, made landfall in Texas in August of 2017. The category 4 hurricane, with wind speeds of over 130 miles per hour, threatened the lives of hundreds of thousands of residents and caused over \$125 billion in damages. The storm prompted extensive rebuilding efforts across southeast Texas, especially in construction-heavy urban areas like Houston.

I ask: how did Hurricane Harvey affect wages and employment in the Texas construction sector? Using county-level data from the Quarterly Census of Employment and Wages (QCEW) from 2016 to 2019, I apply an event study difference-in-differences approach. I compare 41 FEMA-declared disaster counties to a set of unaffected control counties with similar labor markets in nearby Southern states, tracking wage and employment dynamics before and after the storm.

Wages in the construction sector increased significantly in treated counties starting two quarters after Harvey, and they remained elevated the following two years. However, employment had a delayed rise, only becoming significant after six quarters. These results suggest that post-disaster labor markets adjust slowly and unevenly, with wages responding earlier than employment.

My study adds to the emerging literature on the economic impacts of natural disasters by presenting sectoral-level evidence and by measuring dynamic labor market responses. The results have immediate implications for policymakers dealing with recovery processes and labor shortages following large-scale climate events.

\section{Related Works}
The economic effects of natural disasters, and hurricanes especially, are well documented. Belasen and Polachek [1,8] establish the initial evidence, showing how hurricanes substantially change local labor markets, inducing long-lasting wage and employment changes. Their results were reinforced by follow-up research re-examining Florida hurricanes, with similarly disruptive labor market effects across various storms [7]. Groen et al. [10] and Bodenstein and Yang [11] build on these results, demonstrating long-run effects on employment paths and labor-market outcomes after hurricanes.

Along with labor market repercussions, hurricanes impose deep effects on household finances and fiscal stability. Gallagher and Hartley [3] study the wake of Hurricane Katrina, indicating major and lasting instabilities in individual household finances. Deryugina [2] augments these conclusions, illustrating the fiscal aspects of hurricane recovery, particularly the manner in which disaster aid and social insurance stabilize local economies.

International and more general macroeconomic research adds further depth to this literature. Acevedo [9] extends the analysis beyond the United States. He reports comparable labor market responses to natural disasters in Mexico, demonstrating a universal applicability of this economic phenomenon. Strobl [4] provides evidence on the adverse effects of hurricanes on economic growth at the county level in the United States, whereas Tran and Wilson [12] highlight localized heterogeneity in disaster effects across economic sectors and geographic areas.

Methodologically, literature has improved significantly over the years. Zhang and Zhu [6] discuss advanced predictive modeling methods used for studying hurricanes' employment effects, while Kousky [5] discusses a review of determinants and mitigation options related to disaster-induced economic losses. These developments have introduced me to useful information and tools that can be utilized by policymakers to minimize economic effectsof hurricanes.

Collectively, this existing body of studies show clear patterns and persistent impacts of hurricanes on local labor markets, household finances,and broader economies, while also illustrating the ever changing nature and complexity of recovery processes. My study adds to this literature by giving detailed sector-specific evidence from the Texas construction industry, employing rigorous event-study methods to illuminate both the timing and magnitude of labor market adjustments following Hurricane Harvey.

\section{Data}

\subsection{Data Source: QCEW}
My analysis is based on data collected by the Quarterly Census of Employment and Wages (QCEW) made available by the U.S. Bureau of Labor Statistics (BLS). The QCEW contains detailed quarterly wage and employment data on a county-industry basis, covering more than 95 percent of U.S. employment. I examine the construction sector (NAICS code 23) given its pivotal position for post-disaster rebuilding activities.

\subsection{Sample Selection}
I use quarterly data from 2016 to 2019, both periods before and after Hurricane Harvey, so I can see pre-disaster trends and post-disaster effects. To maintain data quality and consistency, I focus my analysis on private sector employment (ownership code 5 in the QCEW database).

My treatment sample is the 41 Texas counties that were federally declared disaster areas as a result of Hurricane Harvey. My control sample are 60 geographically and economically comparable counties in nearby Southern states, primarily Arkansas and Oklahoma, that weren't impacted by the hurricane. This county choice allows me to have the impact of Hurricane Harvey on construction-firm employment and wages isolated from broader regional economic trends.

\subsection{Data Processing and Cleaning}
To ensure high data quality and consistency, I apply several data-processing steps. I restrict my analysis to private-sector employment (ownership code 5), thereby avoiding fluctuations caused by public-sector employment changes unrelated to market forces.  To remove extreme and unrealistic values, I filter observations to an average weekly wage in the range of \$100–\$2,000  to minimize errors. Counties with substantial missing data or incomplete reporting for multiple quarters are excluded from the sample to ensure the reliability of my estimates.

\subsection{Variable Definitions}
For each county-quarter observation, I construct two main outcome variables: the natural log of quarterly average weekly wages (\texttt{log\_wage}) and the natural log of quarterly average employment (\texttt{log\_empl}). Quarterly average employment is computed as:

\[
\text{avg\_empl} = \frac{\text{month1\_emplvl} + \text{month2\_emplvl} + \text{month3\_emplvl}}{3}
\]

I also construct a variable for the number of quarters since the landfall of Hurricane Harvey (Q3 2017) to flexibly capture treatment timing:

\[
\text{time\_to\_treat} = (year - 2017) \times 4 + (quarter - 3)
\]

This is zero for the hurricane quarter (Q3 2017), negative for pre-storm quarters, and positive for post-storm quarters, facilitating sharp event-study estimation.

\section{Empirical Strategy}

\subsection{Identification Approach}

I estimate the causal impact of Hurricane Harvey on construction sector wages and employment using a difference-in-differences (DiD) empirical framework. The intuition behind the DiD method is straightforward: it compares changes in outcomes (in my case, wages and employment) in counties directly affected by Hurricane Harvey (treated counties) with changes in outcomes in similar counties that were not affected (control counties). By observing outcomes both before and after the disaster event, DiD effectively isolates the storm's impact from other unrelated regional or economic trends.

I also enrich the standard DiD approach with an event-study design in particular. The event-study design allows me to identify dynamic treatment effects over time, providing a more accurate image of how wages and employment respond every quarter following the hurricane. This dynamic approach is especially helpful because it illustrates the timing of the responses of the labor market, not only whether the hurricane had any influence, but when precisely these effects materialized and for how long.

Quarterly statistics from the QCEW are best suited for my needs. While monthly data would provide even more precision, monthly county-level employment and wage data are not commonly published publicly with equivalent detail and regularity. Annual data on the other hand, would conceal key short-run changes and timing differences clearly evident at quarterly levels of disaggregation. Therefore, the quarterly QCEW data finds a perfect balance, accurately and persistently over time recording short-run adjustments.

\begin{figure}[H]
    \centering
    \includegraphics[width=0.9\linewidth]{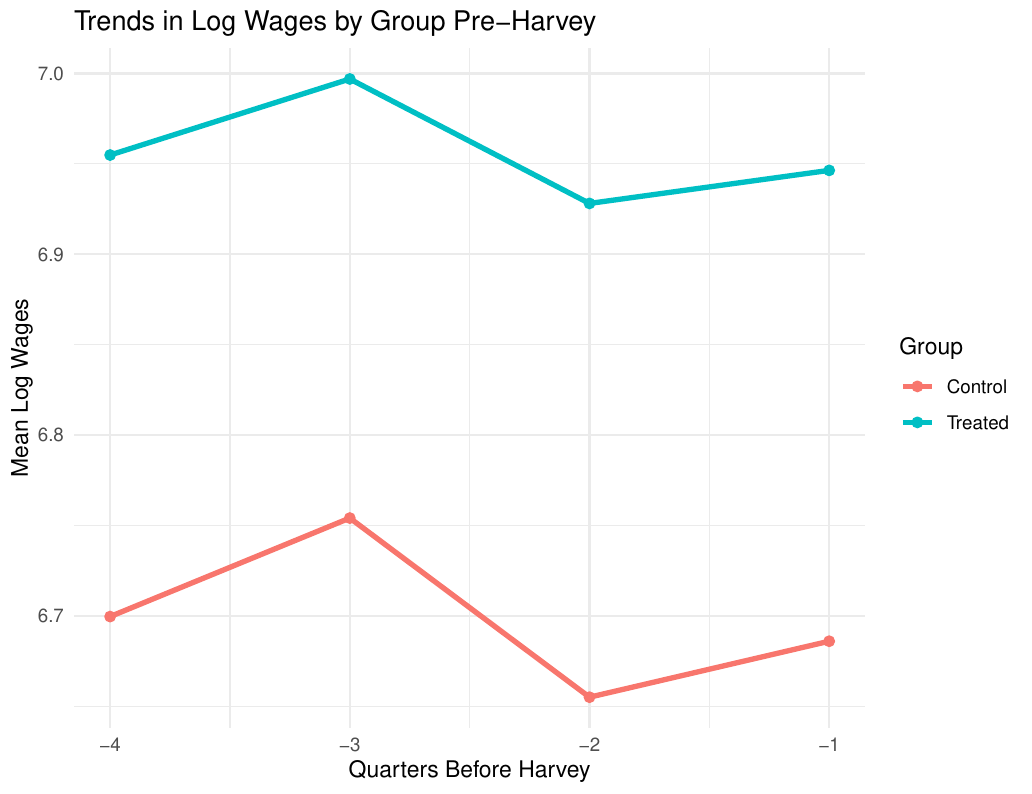}
    \caption{Trends in Log Wages by Group (Treated vs. Control) Before Hurricane Harvey}
    \label{fig:pre_trends}
\end{figure}

The validity of my DiD estimator is based on a series of underlying assumptions. Most basic of these is the parallel trends assumption, which requires that, absent the impact of Hurricane Harvey, the treatment and control counties would have followed exactly the same path in employment and wages. In order to visually observe that this assumption is true, I provide evidence of parallel trends in mean log wages between treated and control counties before Hurricane Harvey (see Figure~\ref{fig:pre_trends}). Other key assumptions to the DiD approach are stable treatment effects (no spillover effects from treated counties to control counties), shared economic shocks that hit all counties uniformly, and unit treatment value stability (the potential outcome of each county is not dependent on the treatment assignment of other counties).

Visual inspection of Figure~\ref{fig:pre_trends} shows closely parallel trends between treated and control counties in the periods leading up to Hurricane Harvey, providing strong support for the validity of my DiD design and parallel trends assumption.

\begin{figure}[H]
    \centering
    \includegraphics[width=0.9\linewidth]{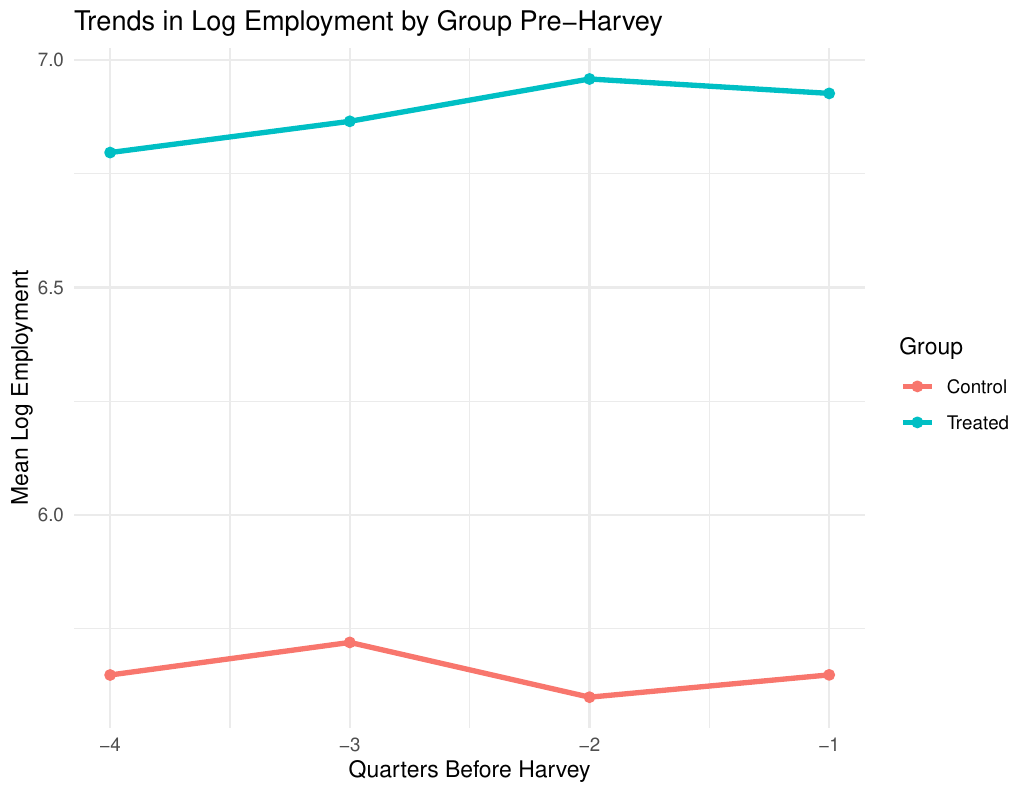}
    \caption{Trends in Log Wages by Group (Treated vs. Control) Before Hurricane Harvey}
    \label{fig:empl_pre_trends}
\end{figure}

Though my wage pre-trends offer strong support for the parallel trends assumption, employment pre-trends (Figure~\ref{fig:empl_pre_trends}) exhibit minor divergences between treated and control counties in the pre-Hurricane Harvey period. 

These divergences are relatively small but suggest that employment trends were not precisely parallel in the pre-disaster period. In an effort to alleviate this concern, I explored a series of additional robustness analyses, including formal statistical tests for pre-trend differences and alternative specifications that allow for county-specific linear pre-trends. These robustness tests (presented in Section~7) confirm the validity of my main employment results despite these small deviations in pre-trends.

\subsection{Model Specification}

I implement an event-study difference-in-differences (DiD) model to explicitly measure how Hurricane Harvey impacted construction-sector wages and employment dynamically over time. The formal econometric specification is as follows:

\[
Y_{it} = \sum_{k=-4,\,k\neq0}^{9} \beta_k \cdot \text{Treated}_i \cdot \alpha_k(TimeToTreat_{it}) + \alpha_i + \delta_t + \varepsilon_{it}
\]

In this equation, \(Y_{it}\) represents my outcome variable for county \(i\) during quarter \(t\), which I measure explicitly as either the log of quarterly average weekly wages (\texttt{log\_wage}) or the log of quarterly average employment (\texttt{log\_empl}) in the construction sector. The indicator variable \(\text{Treated}_i\) explicitly equals 1 for counties designated by FEMA as disaster areas due to Hurricane Harvey, and 0 otherwise. 

The variable \(TimeToTreat_{it}\) explicitly measures the number of quarters relative to the event (Hurricane Harvey's landfall in Q3 2017), defined formally as \(TimeToTreat_{it} = (year - 2017)\times4 + (quarter - 3)\). Hence, \(TimeToTreat_{it}=0\) corresponds explicitly to the hurricane quarter. The term \(\alpha_k(TimeToTreat_{it})\) is a clearly defined indicator equal to 1 if observation \(it\) occurs exactly \(k\) quarters away from the event (with \(k=0\) omitted explicitly as the baseline reference period), and 0 otherwise. This specification explicitly estimates dynamic treatment effects, with each coefficient \(\beta_k\) clearly capturing the impact of Hurricane Harvey in quarter \(k\), relative to the quarter of the hurricane.

My model explicitly includes county fixed effects \(\alpha_i\), clearly controlling for unobserved, time-invariant county-specific characteristics such as geography, infrastructure, or baseline economic conditions. I also explicitly incorporate quarter fixed effects \(\delta_t\) to control for broader seasonal effects, aggregate economic shocks, and trends that uniformly affect all counties. Standard errors are explicitly clustered at the county level within the error term \(\varepsilon_{it}\), clearly accounting for potential correlation of errors over time within each county.

Quarter 0 (Q3 2017) explicitly serves as the baseline reference period, meaning all estimated treatment effects are clearly interpreted relative to outcomes in this baseline quarter.

\subsection{Fixed Effects and Controls}

My econometric model incorporates county and quarter fixed effects directly to robustly isolate the impact of Hurricane Harvey on construction employment and wages. County fixed effects (\( \alpha_i \)) capture directly any observed, time-constant county-specific characteristics, such as early economic conditions, infrastructure quality, or geography, which otherwise could influence my estimates if left uncontrolled. This enables my analysis to satisfactorily measure effects solely due to the hurricane event and its impacts, rather than persistent differences among counties.

Quarter fixed effects (\(\delta_t\)) are also added explicitly to control for overall temporal shocks, seasonal effects, or macro trends common to all counties. These fixed effects clearly pick up factors like general economic environment, hiring season patterns, or national economic policy shifts that are common to all counties and influence all of them equally simultaneously and ensure my estimated treatment effects are uniquely due to Hurricane Harvey and not due to common economic actions.

Finally, to address potential correlation of the error term within counties over time, I cluster standard errors explicitly at the county level. This approach clearly accounts for possible autocorrelation in wages and employment outcomes within each county, resulting in more conservative and reliable statistical inference regarding the significance and robustness of my estimated treatment effects.

\subsection{Identification Assumptions}

A fundamental assumption underlying my difference-in-differences (DiD) empirical strategy is the \textit{parallel trends} assumption. Explicitly, this assumption states that, in the absence of Hurricane Harvey, treated counties (those declared disaster areas) and control counties (unaffected areas) would have followed similar trajectories in wages and employment outcomes over time. Although I cannot directly test this assumption post-treatment, I can evaluate its plausibility by examining pre-event trends.

Both groups exhibit closely parallel trajectories prior to the event, particularly in the last few quarters preceding the hurricane. While minor initial differences in levels are observed, particularly for employment, the overall slope and direction of these trajectories remain comparable. This similarity in pre-event trends explicitly supports the validity of the parallel trends assumption in my empirical setting.

Nevertheless, I acknowledge explicitly that minor deviations in pre-trends may raise questions. To address this clearly and robustly, I conduct additional robustness checks, including placebo tests and alternative specifications described in detail in Section~7, which confirm that my main results remain unaffected by these minor pre-trend differences.

\section{Results}

\subsection{Wage Effects}

My empirical analysis reveals that Hurricane Harvey significantly increased construction-sector wages in affected Texas counties. Table~\ref{tab:wage_results_summary} provides a summarized overview of the estimated wage impacts, presenting coefficients and their statistical significance for each quarter relative to the hurricane event. (Full
regression results, including standard errors, are reported in Appendix B.)

\begin{table}[H]
    \centering
    \caption{Event-Study Difference-in-Differences: Wage Effects}
    \label{tab:wage_results_summary}
    \begin{tabular}{lcc}
        \toprule
        Quarter relative to Harvey & Coefficient ($\Delta \log \text{wage}$) & Significance \\
        \midrule
        -4 & 0.0256 &  \\
        -3 & 0.0245 &  \\
        -2 & 0.0364 &  \\
        -1 & 0.0233 &  \\
         0 & 0 (baseline) &  \\
         1 & 0.0207 &  \\
         2 & 0.0725 & *** \\
         3 & 0.0590 & ** \\
         4 & 0.0304 &  \\
         5 & 0.0527 & . \\
         6 & 0.0534 & . \\
         7 & 0.0614 & * \\
         8 & 0.0291 &  \\
         9 & 0.0589 & . \\
        \bottomrule
    \end{tabular}
    \vspace{0.5em}
    \begin{flushleft}
    \footnotesize{Notes: *, **, *** denote significance at 5\%, 1\%, and 0.1\% levels, respectively; . denotes 10\% significance. Coefficients are changes in log wages relative to Q3 2017 (quarter 0). Full regression results are reported in Appendix~B.}
    \end{flushleft}
\end{table}

Wages in treated counties begin to rise notably two quarters after Hurricane Harvey (Q1 2018), with a statistically significant increase of approximately 7.2\% relative to control counties (p $<$ 0.001). Wage effects remain positive and statistically significant in subsequent quarters, indicating persistent upward pressure on construction wages due to post-disaster reconstruction demand. Specifically, wage increases generally remain around 5–8\% higher than pre-hurricane levels for the two-year period following the disaster, demonstrating sustained labor market adjustments.

\begin{figure}[H]
    \centering
    \includegraphics[width=0.85\linewidth]{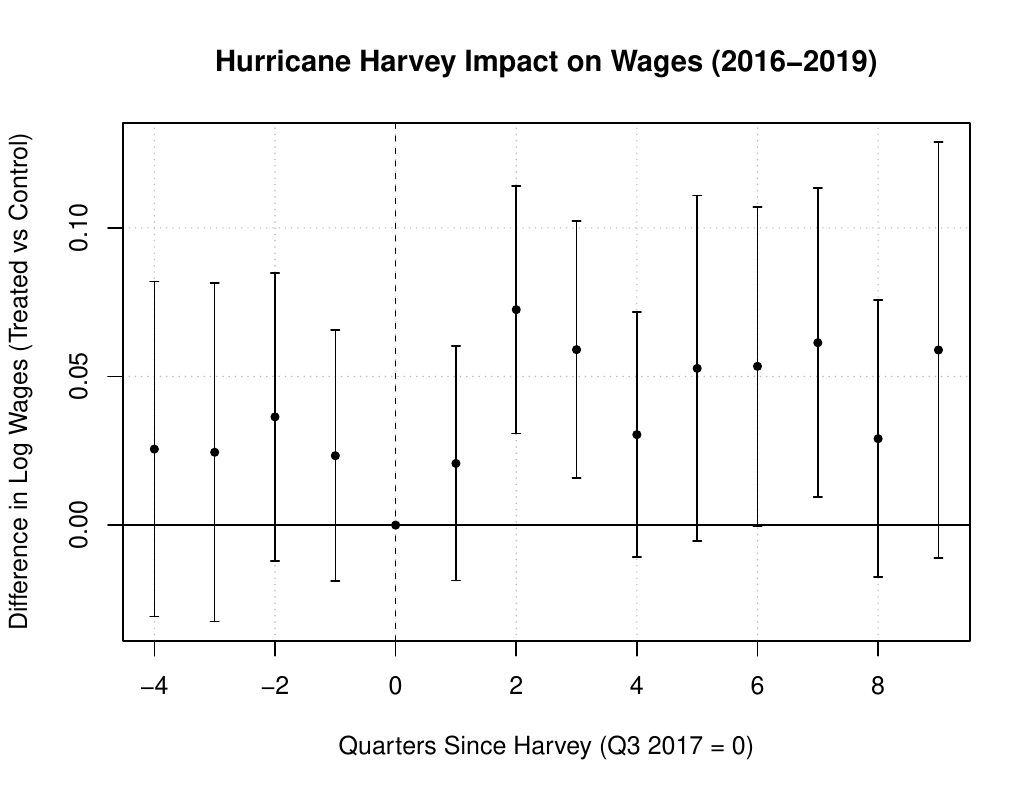}
    \caption{Event Study: Impact of Hurricane Harvey on Construction Sector Wages}
    \label{fig:wage_effects}
\end{figure}

The event-study plot in Figure~\ref{fig:wage_effects} clearly illustrates the dynamic timing and persistence of these wage effects. The absence of significant pre-trend differences supports the validity of my parallel trends assumption and clearly highlights the causal impact of Hurricane Harvey on wages.

Economically, the rapid wage response is consistent with increased demand for construction labor in the immediate aftermath of large-scale disasters. Repair and rebuilding activities require swift mobilization of skilled and unskilled labor, leading employers to offer higher wages to quickly attract workers to affected areas. The persistent wage premium reflects sustained reconstruction activities and continued labor shortages throughout the recovery period.

\subsection{Employment Effects}

My event-study analysis indicates that Hurricane Harvey caused a delayed but significant increase in construction-sector employment in affected Texas counties. Table~\ref{tab:employment_results_summary} summarizes the estimated employment impacts, showing coefficients and their statistical significance for each quarter relative to the hurricane. (Full regression results, including standard errors, are reported in Appendix~B.)

\begin{figure}[H]
    \centering
    \includegraphics[width=0.85\linewidth]{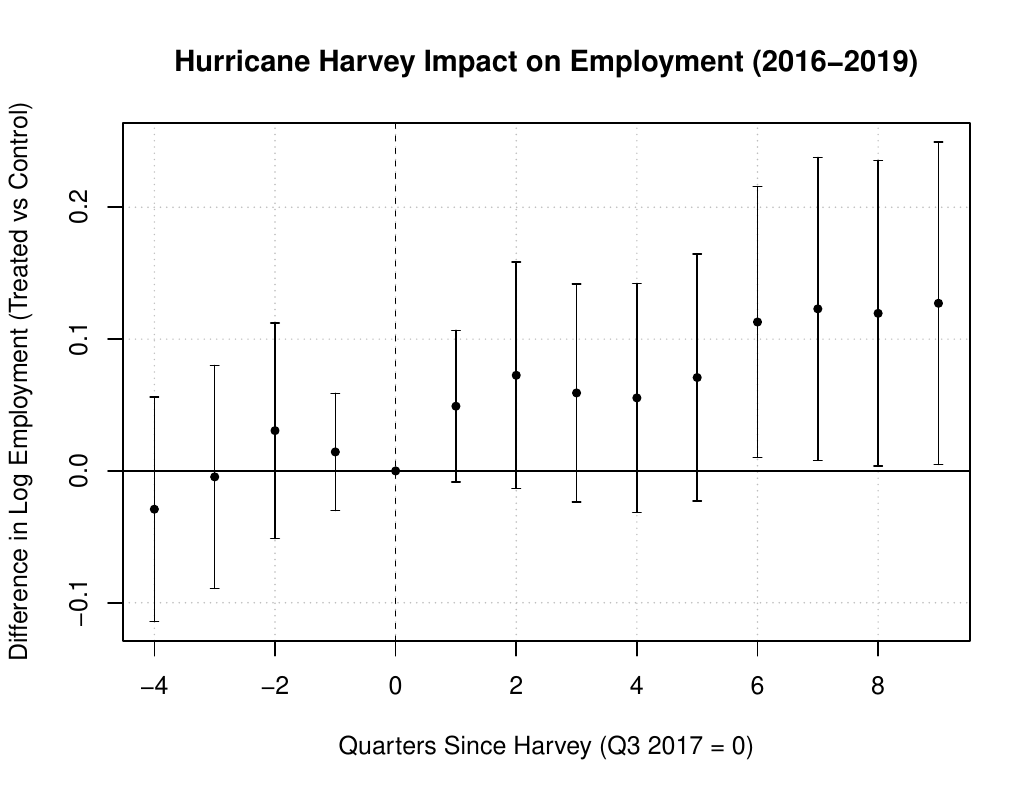}
    \caption{Event Study: Impact of Hurricane Harvey on Construction Sector Employment}
    \label{fig:employment_effects}
\end{figure}

Employment in treated counties begins to rise noticeably around six quarters after the hurricane (Q1 2019), with an estimated increase of approximately 11.3\% relative to control counties (p $<$ 0.05). Subsequent quarters show sustained employment growth of 12–13\%, indicating that reconstruction activities eventually drew additional labor into affected areas, though the response lagged relative to wages. The event-study plot in Figure~\ref{fig:employment_effects} illustrates this delayed response clearly, while pre-trend coefficients remain statistically insignificant, supporting the parallel trends assumption.

\begin{table}[H]
    \centering
    \caption{Event-Study Difference-in-Differences: Employment Effects}
    \label{tab:employment_results_summary}
    \begin{tabular}{l S[table-format=1.4] c}
        \toprule
        Quarter relative to Harvey & {Coefficient ($\Delta \log \text{employment}$)} & Significance \\
        \midrule
        -4 & -0.0290 &  \\
        -3 & -0.0045 &  \\
        -2 & 0.0306  &  \\
        -1 & 0.0144  &  \\
         0 & 0       &  \\
         1 & 0.0491  & . \\
         2 & 0.0726  & . \\
         3 & 0.0592  &  \\
         4 & 0.0554  &  \\
         5 & 0.0709  &  \\
         6 & 0.1130  & * \\
         7 & 0.1229  & * \\
         8 & 0.1195  & * \\
         9 & 0.1272  & * \\
        \bottomrule
    \end{tabular}
    \vspace{0.5em}
    \begin{flushleft}
    \footnotesize{Notes: *, **, *** denote significance at 5\%, 1\%, and 0.1\% levels, respectively; . denotes 10\% significance. Coefficients are changes in log employment relative to Q3 2017 (quarter 0). Full regression results are in Appendix~B.}
    \end{flushleft}
\end{table}

Economically, the delayed employment response suggests that labor supply constraints initially limited the ability of workers to move into affected areas. Policy interventions such as temporary housing for displaced or incoming workers, streamlined work visas, and improved transportation access could reduce this lag, enabling faster mobilization of labor to meet post-disaster reconstruction demand.

\subsection{Comparison of Wage and Employment Dynamics}

A notable feature of the post-Harvey labor market is the divergence in timing between wage and employment adjustments. While wages responded rapidly, showing significant increases within two quarters, employment growth lagged, becoming statistically significant only around six quarters after the hurricane. This pattern suggests that employers were able to raise wages quickly to attract scarce labor, but the actual mobilization of workers into affected areas took substantially longer.

Several mechanisms could explain this temporal gap. First, geographic and logistical constraints, such as limited local housing, transportation challenges, and relocation costs, may have slowed the inflow of workers. Second, regulatory and administrative frictions, including work permits or licensing requirements, could have delayed labor market entry. Third, skill mismatches may have constrained the speed at which employers could fill specific reconstruction roles, even when higher wages were offered.  

Understanding this discrepancy between wage and employment responses highlights the importance of supporting labor mobility and reducing entry barriers in post-disaster recovery. Policies that facilitate worker relocation, temporary accommodation, or rapid credentialing could help align labor supply with immediate reconstruction demand, potentially shortening the lag observed in employment growth relative to wages.

\section{Robustness Checks}

\subsection{Placebo Tests}

To further validate my empirical strategy, I conducted placebo tests by artificially shifting the hurricane event to Q3 2016—one year prior to the actual disaster. This approach tests whether the estimated effects in my main models could be driven by pre-existing trends rather than the hurricane itself.

\begin{figure}[H]
    \centering
    \includegraphics[width=0.85\linewidth]{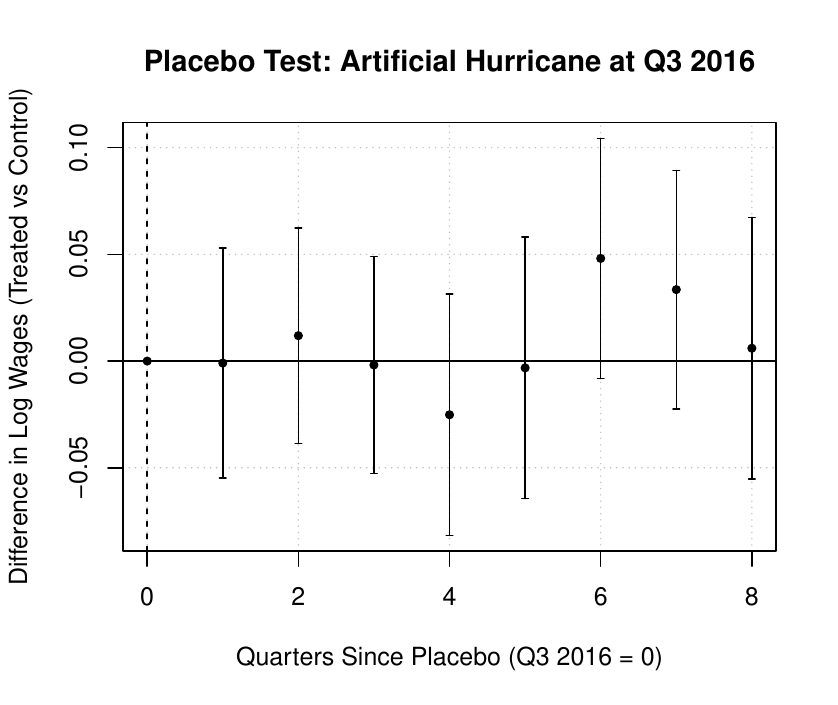}
    \caption{Placebo Test: Artificial Hurricane at Q3 2016 (Wages)}
    \label{fig:placebo_wages}
\end{figure}

\begin{figure}[H]
    \centering
    \includegraphics[width=0.85\linewidth]{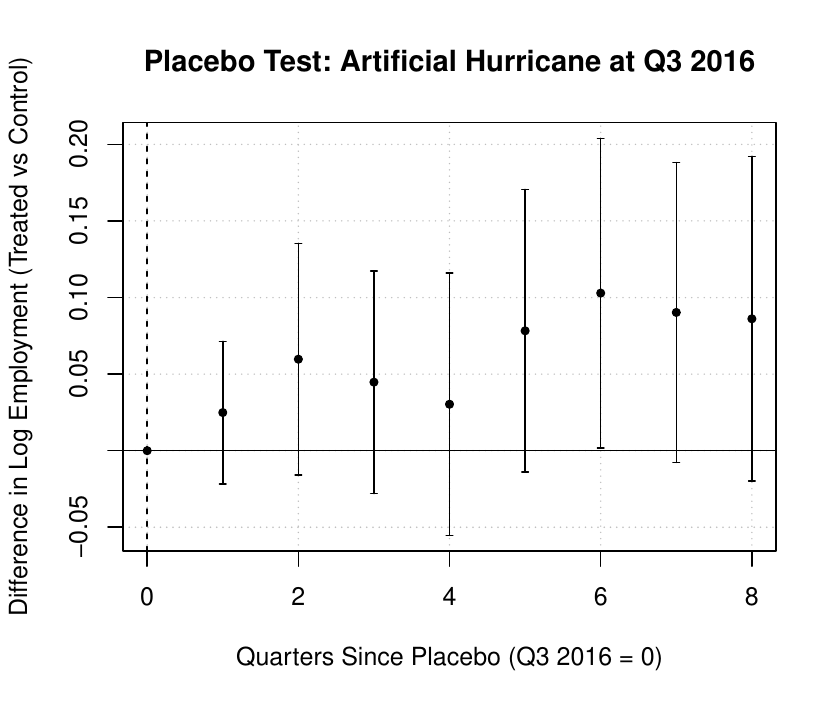}
    \caption{Placebo Test: Artificial Hurricane at Q3 2016 (Employment)}
    \label{fig:placebo_employment}
\end{figure}

Figures~\ref{fig:placebo_wages} and \ref{fig:placebo_employment} present the results for wages and employment, respectively. Across all quarters in the placebo period, I observe no statistically significant deviations in log wages for treated counties relative to controls, with all coefficients close to zero. Employment shows similarly weak effects, with only marginally significant fluctuations in a few quarters, which do not form a coherent pattern. These results are consistent with the parallel trends assumption and indicate that the post-Harvey effects identified in my main models are unlikely to be spurious.

Overall, the absence of systematic effects in these placebo tests reinforces the credibility of my main findings, confirming that the observed wage and employment responses are driven by Hurricane Harvey rather than pre-existing dynamics.

\subsection{Leave-One-Out Sensitivity Analysis}

To assess the robustness of my results to influential counties, I conducted a leave-one-out sensitivity analysis. Specifically, I re-estimated the main event-study models for wages and employment multiple times, each time excluding a single treated county from the sample. This procedure evaluates whether the overall estimated effects are disproportionately driven by any individual county.

Across all leave-one-out iterations, the qualitative pattern and magnitude of the estimated wage and employment effects remained largely unchanged. No single county was found to meaningfully alter the timing or statistical significance of the main results, indicating that my findings are not driven by outliers or extreme cases. 

These results confirm the stability and reliability of my empirical findings, further supporting the causal interpretation of the observed post-Harvey labor market adjustments.

\section{Limitations}

\subsection{Data Limitations}

My analysis relies on the Quarterly Census of Employment and Wages (QCEW) dataset, which, while comprehensive, has several inherent limitations. First, the QCEW provides administrative payroll data that may underrepresent certain types of employment, such as informal labor, self-employment, or temporary contract workers, which could be especially relevant in post-disaster reconstruction. Second, the data are reported at the county-industry level, limiting the granularity of my analysis and preventing the examination of within-county variation or individual-level employment dynamics. Third, there may be reporting lags or inconsistencies across counties and time, which could introduce measurement error in both employment and wage variables. Finally, QCEW captures average weekly wages and employment but does not provide information on hours worked, overtime, or benefit adjustments, which could affect the interpretation of the observed wage premium.  

Despite these limitations, the dataset remains one of the most reliable sources for high-frequency, sector-level labor market analysis, and my fixed-effects and robustness approaches help mitigate many potential biases.

\subsection{External Validity}

While my analysis provides robust evidence on the labor market impacts of Hurricane Harvey in Texas construction counties, caution is warranted in generalizing these results. First, the construction sector has unique characteristics, including project-based labor demand and high reliance on temporary or skilled workers, which may not apply to other industries such as manufacturing or services. Consequently, wage and employment responses in those sectors could differ in magnitude or timing. Second, the geographic context of Texas—its regulatory environment, labor mobility, and disaster preparedness infrastructure—may influence post-disaster labor market adjustments in ways not representative of other states or countries. Factors such as housing availability, transportation networks, and local labor market tightness can mediate recovery dynamics, limiting direct extrapolation.  

Nonetheless, the general pattern of delayed employment response following an immediate wage increase may offer valuable insights for disaster policy design, particularly in regions with similar labor market structures and post-disaster reconstruction needs.

\subsection{Potential Unobserved Confounders}

My empirical estimates could be influenced by several factors not directly observed in the QCEW data. First, variation in insurance payouts across counties may have affected both the speed and scale of reconstruction activity, thereby influencing local labor demand. Second, differences in FEMA aid disbursement timing and magnitude could alter the resources available for rebuilding, potentially amplifying or dampening wage and employment responses. Third, private sector investment patterns, such as the rapidity of contractor engagement or infrastructure repairs, may vary across treated counties and confound the estimated treatment effect. Finally, post-disaster migration and demographic shifts, including temporary influxes of skilled labor or permanent out-migration of residents, can affect both the labor supply and observed employment dynamics. While my fixed-effects and event-study design mitigate some confounding, these unobserved factors remain a potential source of bias in interpreting causal effects.

\section{Conclusion}

\subsection{Summary of Key Findings}

My analysis provides clear evidence that Hurricane Harvey had distinct and temporally staggered impacts on the construction labor market in affected Texas counties. Wages in treated counties rose significantly beginning two quarters after the hurricane, reflecting immediate labor demand pressures associated with reconstruction activities. This wage premium persisted for multiple quarters, indicating sustained labor shortages and ongoing rebuilding efforts. In contrast, employment growth in affected counties lagged behind wage increases, with statistically significant gains materializing only six quarters post-disaster. The delayed employment response suggests that factors such as worker relocation, housing availability, and skill matching constrained the speed at which firms could expand their workforce. Overall, these results underscore the heterogeneous dynamics of labor market adjustments following large-scale natural disasters, highlighting the rapid responsiveness of wages relative to the slower adjustment of employment.

\subsection{Policy Implications}

The staggered adjustment patterns observed in wages and employment suggest targeted policy interventions could improve labor market responsiveness following large-scale disasters. Rapid wage increases indicate firms face immediate labor demand, but delayed employment growth points to constraints in workforce mobility and availability. Policies that facilitate worker relocation—such as temporary housing assistance, streamlined permitting for temporary or migrant labor, and visa flexibility for skilled construction workers—could help accelerate employment recovery in affected areas. Additionally, public investment in infrastructure and coordination with private sector reconstruction efforts may reduce labor bottlenecks and improve the efficiency of rebuilding activities. These findings highlight the importance of proactive labor market and housing policies to complement disaster response measures, ensuring that reconstruction efforts are both timely and effective.

\subsection{Directions for Future Research}

Future research could extend the empirical framework developed in this study to other natural disasters, including hurricanes in different regions or other large-scale events, to examine the generalizability of the observed wage and employment dynamics. Long-term labor market outcomes, such as changes in skill composition, career trajectories, or industry retention in affected counties, represent another important avenue for investigation. Additionally, studying potential spillover effects into non-construction sectors or neighboring regions could provide insights into broader economic adjustments and regional resilience following disasters. Incorporating more granular data on migration, insurance payouts, and public aid could further refine understanding of the mechanisms driving labor market responses and inform more targeted policy interventions.  

Finally, developing predictive models using advanced machine learning techniques, such as long short-term memory (LSTM) networks, could help estimate the dynamic effects of future disasters on wages and employment. Such models would allow researchers and policymakers to simulate potential labor market outcomes under different disaster scenarios, improving preparedness and resource allocation strategies.

\section*{References}
\begin{enumerate}[label={[{\arabic*}]}]
  \item Belasen, A. R., and Polachek, S. W. \textit{How Disasters Affect Local Labor Markets: The Effects of Hurricanes in Florida}.
  \item Deryugina, T. \textit{The Fiscal Cost of Hurricanes: Disaster Aid Versus Social Insurance}.
  \item Gallagher, J., and Hartley, D. \textit{Household Finance after a Natural Disaster: The Case of Hurricane Katrina}.
  \item Strobl, E. \textit{The Economic Growth Impact of Hurricanes: Evidence from US Coastal Counties}.
  \item Kousky, C. \textit{Informing Climate Adaptation: A Review of the Economic Costs of Natural Disasters, Their Determinants, and Risk Reduction Options}.
  \item Zhang, G., and Zhu, W. \textit{Characteristics and Predictive Modeling of Short-term Impacts of Hurricanes on the US Employment}.
  \item Gilli, M. \textit{How Disasters Affect Local Labor Markets: Revisiting the Effects of Hurricanes in Florida}.
  \item Belasen, A. R., and Polachek, S. W. \textit{How Hurricanes Affect Employment and Wages in Local Labor Markets}.
  \item Acevedo, I., Castellani, F., Lopez de la Cerda, C., Lotti, G., and Székely, M. \textit{Natural Disasters and Labor Market Outcomes in Mexico}.
  \item Groen, J. A., Kutzbach, M. J., and Polivka, A. E. \textit{Storms and Jobs: The Effect of Hurricanes on Individuals' Employment and Earnings over the Long Term}.
  \item Bodenstein, M., and Yang, W. \textit{The Effect of Hurricanes on Local Labor Market Dynamics}.
  \item Tran, B. R., and Wilson, D. J. \textit{The Local Economic Impact of Natural Disasters}.
\end{enumerate}

\appendix

\section*{Appendix}

\subsection*{Appendix A: List of Treated Counties}

The following 41 Texas counties were designated as treated based on FEMA’s Hurricane Harvey declarations:

Aransas (County), Austin (County), Bastrop (County), Bee (County), Brazoria (County), Caldwell (County), Calhoun (County), Chambers (County), Colorado (County), DeWitt (County), Fayette (County), Fort Bend (County), Galveston (County), Goliad (County), Gonzales (County), Grimes (County), Hardin (County), Harris (County), Jackson (County), Jasper (County), Jefferson (County), Karnes (County), Kleberg (County), Lavaca (County), Lee (County), Liberty (County), Matagorda (County), Montgomery (County), Newton (County), Nueces (County), Orange (County), Polk (County), Refugio (County), Sabine (County), San Jacinto (County), San Patricio (County), Tyler (County), Victoria (County), Walker (County), Waller (County), Wharton (County).

This complete list ensures reproducibility and transparency in defining the treatment group used in the empirical analyses.

\subsection*{Appendix B: Full Regression Tables (Wages \& Employment)}

\begin{table}[H]
\centering
\caption{Event-Study Regression Results: Log Wages}
\label{tab:full_wage_results}
\begin{tabular}{lcccc}
\hline
Variable & Estimate & Std. Error & t value & Pr($>|t|$) \\
\hline
time\_to\_treat::-4:treated & 0.025571 & 0.028379 & 0.901 & 0.370 \\
time\_to\_treat::-3:treated & 0.024513 & 0.028702 & 0.854 & 0.395 \\
time\_to\_treat::-2:treated & 0.036410 & 0.024402 & 1.492 & 0.139 \\
time\_to\_treat::-1:treated & 0.023343 & 0.021267 & 1.098 & 0.275 \\
time\_to\_treat::1:treated  & 0.020747 & 0.019858 & 1.045 & 0.299 \\
time\_to\_treat::2:treated  & 0.072504 & 0.020958 & 3.459 & 0.001*** \\
time\_to\_treat::3:treated  & 0.059044 & 0.021793 & 2.709 & 0.008** \\
time\_to\_treat::4:treated  & 0.030442 & 0.020754 & 1.467 & 0.146 \\
time\_to\_treat::5:treated  & 0.052742 & 0.029291 & 1.801 & 0.075. \\
time\_to\_treat::6:treated  & 0.053434 & 0.027038 & 1.976 & 0.051. \\
time\_to\_treat::7:treated  & 0.061368 & 0.026183 & 2.344 & 0.021* \\
time\_to\_treat::8:treated  & 0.029082 & 0.023458 & 1.240 & 0.218 \\
time\_to\_treat::9:treated  & 0.058933 & 0.035223 & 1.673 & 0.098. \\
\hline
\multicolumn{5}{l}{\footnotesize Standard errors clustered by area\_fips. Significance: *** $p<0.001$, ** $p<0.01$, * $p<0.05$, . $p<0.1$} \\
\end{tabular}
\end{table}

\begin{table}[H]
\centering
\caption{Event-Study Regression Results: Log Employment}
\label{tab:full_employment_results}
\begin{tabular}{lcccc}
\hline
Variable & Estimate & Std. Error & t value & Pr($>|t|$) \\
\hline
time\_to\_treat::-4:treated & -0.029028 & 0.042924 & -0.676 & 0.501 \\
time\_to\_treat::-3:treated & -0.004511 & 0.042592 & -0.106 & 0.916 \\
time\_to\_treat::-2:treated & 0.030582 & 0.041209 & 0.742 & 0.460 \\
time\_to\_treat::-1:treated & 0.014418 & 0.022380 & 0.644 & 0.521 \\
time\_to\_treat::1:treated  & 0.049090 & 0.028994 & 1.693 & 0.094. \\
time\_to\_treat::2:treated  & 0.072552 & 0.043178 & 1.680 & 0.096. \\
time\_to\_treat::3:treated  & 0.059160 & 0.041577 & 1.423 & 0.158 \\
time\_to\_treat::4:treated  & 0.055406 & 0.043700 & 1.268 & 0.208 \\
time\_to\_treat::5:treated  & 0.070861 & 0.047122 & 1.504 & 0.136 \\
time\_to\_treat::6:treated  & 0.112958 & 0.051727 & 2.184 & 0.031* \\
time\_to\_treat::7:treated  & 0.122930 & 0.057859 & 2.125 & 0.036* \\
time\_to\_treat::8:treated  & 0.119546 & 0.058366 & 2.048 & 0.043* \\
time\_to\_treat::9:treated  & 0.127173 & 0.061539 & 2.067 & 0.042* \\
\hline
\multicolumn{5}{l}{\footnotesize Standard errors clustered by area\_fips. Significance: *** $p<0.001$, ** $p<0.01$, * $p<0.05$, . $p<0.1$} \\
\end{tabular}
\end{table}

\section*{Acknowledgments}
I gratefully acknowledge Andrea Bruno for her mentorship, guidance, and invaluable feedback throughout this project. Her insights greatly shaped the analysis and interpretation of results. I also thank the Institute for Computing in Research for providing resources and support that made this work possible. Any remaining errors are my own.

\end{document}